\documentclass[aps,twocolumn,showpacs,pra,eqsecnum]{revtex4}
\usepackage{bm}
\usepackage{graphicx}
\usepackage{amsbsy}
\usepackage{amsmath}
\usepackage{amsfonts}
\usepackage{amssymb}
\usepackage{amsmath}
\usepackage{epic,eepic}
\usepackage{graphicx}
\usepackage{amsthm}

\newtheorem*{lemA.1}{Lemma A.1}
\newtheorem*{lemA.2}{Lemma A.2}
\newtheorem*{lemA.3}{Lemma A.3}

\newtheorem{theorem}{Theorem}

\newtheorem{problem}{Problem}

\newtheorem{example}{Example}

\begin{document}
\title{On continuous variable quantum algorithms for oracle identification problems}
    \author{Mark R. A. Adcock,$^{1}$ Peter H\o yer,$^{1,2}$ and Barry C. Sanders$^{1}$
   }
 \affiliation{$^{1}$Institute for Quantum Information Science,
    University of Calgary, Calgary, Alberta, Canada, T2N 1N4. Email: \texttt{mkadcock@qis.ucalgary.ca}\\
 $^{2}$Department of Computer Science,
University of Calgary, 2500 University Drive N.W., Calgary,
Alberta, Canada, T2N 1N4. Email: \texttt{hoyer@ucalgary.ca}. }

\begin{abstract}

We establish a framework for oracle identification problems in the
continuous variable setting, where the stated problem necessarily
is the same as in the discrete variable case, and continuous
variables are manifested through a continuous representation in an
infinite-dimensional Hilbert space. We apply this formalism to the
Deutsch-Jozsa problem and show that, due to an uncertainty
relation between the continuous representation and its
Fourier-transform dual representation, the corresponding
Deutsch-Jozsa algorithm is probabilistic hence forbids an
exponential speed-up, contrary to a previous claim in the
literature.

\end{abstract}

 \maketitle

\noindent December 18, 2008
\section{Introduction}
\label{sec:intro}

Quantum information protocols have been demonstrated
experimentally in both the discrete-variable (DV) and so-called
continuous-variable (CV) settings. DV quantum information
protocols employ qubits~\cite{NC00} and qudits~\cite{GKP01}, and
CV quantum information protocols regard continuously parameterized
canonical position states as the logical elements analogous to
qubits for the DV case~\cite{BP03}. CV quantum information is
experimentally appealing because sophisticated squeezed light
experiments have led to claims of successful quantum information
protocols such as teleportation~\cite{FSBFK98}, key
distribution~\cite{GG02}, and memory~\cite{AYANTFK07,AFKLL08}, but
the theoretical status of CV quantum information is challenged by
unresolved issues concerning quantum error correction~\cite{Br98},
non-distillability~\cite{EP02}, no-go theorems for quantum
computation~\cite{BSBN02,BS02}, and the absence of full security
proofs for key distribution.

CV information processing has also been studied for classical
models, including the now named Blum-Shub-Smale machine
\cite{BSS89} and continuous Turing machines \cite{MP}. These
models are of background relevance to the research into CV quantum
information and are referenced here for contextual purposes.

In this paper, we establish a sound theoretical framework for
studying quantum algorithms and apply this framework to study the
CV analogue of the early DV quantum algorithm, known  as the
Deutsch-Jozsa (DJ) algorithm~\cite{DJ92,DD85,CEMM98}. The problem
solved by DJ algorithm is the following.
\begin{problem}
\label{problem:balancedorconstant}
    Given a function $f:\{0,1\}^n\rightarrow\{0,1\}$
    that is promised to be either \emph{constant} ($f$ takes the same value everywhere)
    or \emph{balanced} ($f$ takes the value 0 on exactly half the inputs),
    determine whether $f$ is constant or balanced.
\end{problem}
The best classical algorithm requires $2^{n-1}+1$ evaluations in
the worst case. If error is tolerated, for any integer $m \geq 2$,
to achieve an error of at most $2^{-m}$, any probabilistic
algorithm requires a number of evaluations that is at least of
order~$m$ ~\cite{Ba05}. If the function is accessible on a quantum
computer as a quantum oracle, then the DJ algorithm is exact and
requires just one evaluation to solve the problem.

Our focus here is on the CV analogue of the DJ algorithm, and we
are inspired by the Braunstein and Pati formulation~\cite{BP02} of
the CV DJ algorithm; however, our work differs from theirs in that
ours relies only on logical states that are elements in the
Hilbert Space and thus provides a strict CV version of the DJ
problem. We introduce a particular model for the computation of
the DJ problem in a CV setting. Within the constraints of this
model, our analysis shows that the CV DJ algorithm is necessarily
probabilistic and its performance must therefore be compared to
the classical case where bounded error is tolerated and not to the
classical deterministic case.

We choose the DJ algorithm for the following reasons. Two types of
quantum algorithms dominate the field, those that implement a
version of the hidden subgroup problem and those that use a
version of Grover's search algorithm~\cite{NC00,Gro96}. An early
example of the former is the Deutsch-Jozsa (DJ)
algorithm~\cite{DJ92}, which is amongst the oracle class of
problems~\cite{AIKRY04} that have been important in demonstrations
of quantum speed-ups. Finally the CV DJ algorithm has a head start
in the work of Braunstein and Pati so our analysis can build on
their concepts~\cite{BP02}.

Our paper is presented as follows. In Sec.~\ref{sec:background},
we review the DJ algorithm. Although this algorithm is well known,
our review serves as a foundation for careful construction of the
CV version of this algorithm. Furthermore we compare the DJ
algorithm's performance against both deterministic and
probabilistic strategies, especially because the CV case can only
be properly compared against probabilistic strategies because the
CV DJ algorithm can never be deterministic. Our description of the
DV DJ algorithm comprises three steps so that these steps can be
discussed separately during the construction of the CV analogue.

Our approach emphasizes a recasting of the DV DJ algorithm in that
we do not need the target qubit. This approach leads to an easier
adaptation to the CV case. In Sec.~\ref{sec:background}, we review
the formalism of rigged Hilbert spaces (RHS)~\cite{dlM05} since
our CV algorithm, as well as any other CV quantum algorithms, must
work in a RHS.  This will have implications when we discuss the
limitations of error inherent in our CV DJ algorithm in
Sec.~\ref{sec:CVrep} and in Sec.~\ref{sec:bounding}.

In Sec.~\ref{sec:CVrep}, we adapt the DJ problem to the CV case
and develop the CV DJ algorithm through the same three fundamental
steps of the algorithm. We pay particular attention to the
challenge of encoding a finite $N$-bit string into functions over
the real numbers. Overcoming this challenge enables us to
recognize that perfect encoding results in the inability to
determine if the encoding is of a constant string or balanced in a
single execution of the algorithm. We show that this probabilistic
nature of the algorithm is the result of an uncertainty relation
between the continuous representation and its Fourier-transform
dual representation.

In Sec.~\ref{sec:bounding}, we determine an upper bound on the
query complexity of the CV DJ algorithm. We note that because the
CV DJ algorithm is shown to be probabilistic, its performance can
only logically be compared to the classical probabilistic
algorithm and not to the classical deterministic algorithm. We
conclude that the formalism presented herein is applicable to a
wide range of oracle identification problems in a CV setting.

\section{Background}
\label{sec:background}

We cast the DJ problem into the class of `oracle identification
problems' in Subsec.~\ref{subsec:oracle}. We then review
deterministic algorithms in Subsec.~\ref{subsec:deterministic} and
probabilistic algorithms in Subsec.~\ref{subsec:randomized}. In
Subsec.~\ref{subsec:quantumDJ}, we analyze an alternative
representation of the quantum DJ algorithm that uses $n$ qubits
instead of the traditional $n+1$ qubits. In
Subsec.~\ref{subsec:CV}, we present a primer on the rigged Hilbert
space and close with a discussion of the concepts required to
transition from discrete variables to continuous variables.

\subsection{The Oracle Identification Problem}
\label{subsec:oracle}

The DJ problem is an identification problem in which we are given
a function from some candidate set ${\mathcal S} = \{f_1, f_2,
\ldots, f_M\}$ of functions.  The candidate set ${\mathcal S} =
{\mathcal S}_0 \cup {\mathcal S}_1$ is the disjoint union of two
collections of functions, and our task is to determine which of
the two collections the function $f$ is drawn from.
\begin{problem}
  \label{problem:oip}
  Consider the set of all functions from $n$ bits to one bit,
  ${\mathcal F} ={ \{f \mid f :\{0,1\}^n \rightarrow \{0,1\}\,\}}$.  Let $S_0$
    and $S_1$ be disjoint subsets of $\mathcal F$.  Given some oracle
    \begin{equation}
    \label{eq:oraclef}
        f: \{0,1\}^n \rightarrow \{0,1\}
    \end{equation}
    with the promise that either $f \in S_0$ or $f \in S_1$, determine
    the index $b$ such that $f \in S_b$.
\end{problem}

For $N=2^n$, we impose lexicographic order on the $N$-bit strings
of $\{0,1\}^n$. We can then specify any function $f_z$ by writing
all its $N$ function values in a list $z \in \{0,1\}^N$ of
length~$N$.  The $i^{\textup{th}}$ bit $z_i$ in the list is 1 if
$f$ takes the value~1 on the $i^{\textup{th}}$ bit-string of
$\{0,1\}^n$. There are $2^N$ functions from $n$ bits to one bit,
and thus our candidate set has cardinality upper bounded by~$M
\leq 2^N$.  In the following, we often write $f_z$ to denote the
function that corresponds to the $N$-bit string~$z$.

We are interested in finding an efficient strategy to identify the
property of whether $f$ belongs to set $S_0$ or to set $S_1$
without necessarily determining $f$ itself.  In the DJ case, the
property we are interested in is whether $f$ is balanced or
constant~\cite{DJ92,DD85,CEMM98}.  The cost of the algorithm is
the number of queries made to the oracle.

With the promise of balanced or constant functions, there are far
fewer than $2^N$ functions.  The number of balanced and the number
of constant functions is readily ascertained from the binomial
theorem applied to power sets.  The strings $z$ of length $N$ that
correspond to the constant functions are the string consisting
only of~1s and the string consisting only of~0s.  There are thus
just two constant functions.  The strings $z$ of length $N$ that
correspond to balanced functions are the strings in which exactly
half of the bits are~0 and half are~1.  There are thus precisely
${N \choose N/2}$ balanced functions.

\subsection{The Classical Deterministic Approach}
\label{subsec:deterministic}

On a classical Turing machine,
Problem~\ref{problem:balancedorconstant} can be solved
deterministically. A deterministic algorithm corresponds to
submitting queries in the form of $n$-bit inputs and obtain the
one-bit output for each query. There are~$N$ unique input strings,
but the promise of balanced versus constant functions implies that
only $N/2+1$ are required to determine whether the given function
is balanced or constant, with certainty.

The reason that fewer than $N/2+1$ queries is insufficient is that only $N/2$
queries may reveal all output bits being the same, suggesting a constant
function, whereas the remaining $N/2$ outputs could all be the opposite of
the first $N/2$ queries.

\subsection{The Classical Probabilistic Approach}
\label{subsec:randomized}

In Subsec.~\ref{subsec:deterministic}, we saw that fewer than $N/2+1$ queries is insufficient
for a deterministic algorithm, but that case seems highly unlikely. More formally,
fewer than $N/2+1$ queries will identify most of the balanced functions as non-constant
in much fewer than $N/2+1$ queries. Here we ask the question about how many queries
are required if we are prepared to tolerate a small number of errors.

In fact a probabilistic algorithm achieves an exponentially small
error of $2^{-m}$ with a number of queries that is only linear
in~$m$~\cite{Ba05}. To understand how a probabilistic algorithm
can help, consider that, although a single query with a random
input provides no information, two queries with two random inputs
can be highly informative. If the output from the second query
differs from the first output, then the function is proved not to
be constant and therefore must be balanced. If, on the other hand,
the second output is the same as the first, then the outcome is
not certain, but the more times the outputs are the same, the more
confident one can be about the function being constant.

\begin{figure}[tbp]
            \begin{center}
            \includegraphics[width=8.5cm]{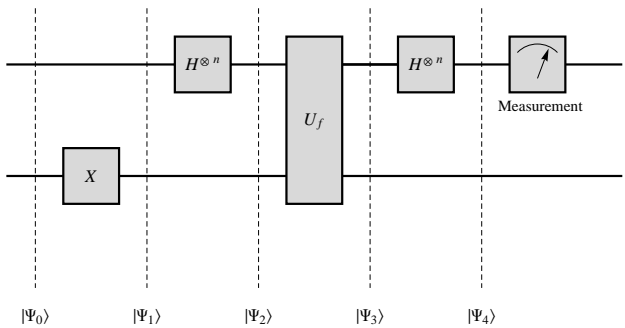}
            \end{center}
            \caption{Quantum circuit implementing the Discrete Variable DJ Algorithm. The upper line represents the
            $n$-qubit ``control'' state, and the lower line represents the 1-qubit ``target'' state.}
\label{fig:tradQDJcircuit}
        \end{figure}

We calculate the probability of successfully determining whether
the given function~$f_z$ is balanced or constant. A lower bound on
the success probability~$\Pr_\checkmark$ for~$m$ queries can be
achieved by examining a sampling-without-replacement strategy,
which is expressed as
\begin{equation}
\label{eq:ClassDJRan}
    \Pr_\checkmark=1-\prod_{j=1}^m\frac{N/2-(j-1)}{N-(j-1)}
        \geq 1-\left(\frac{1}{2}\right)^m.
\end{equation}
Here the equality is calculated assuming sampling without
replacement and shows dependency on~$N$, whereas the inequality in
Eq.~(\ref{eq:ClassDJRan}) is based on sampling with replacement
and is independent of $N$. The failure probability
$1-\Pr_\checkmark$ declines exponentially in~$m$, the number of
queries.

In Subsec.~\ref{subsec:quantumDJ}, we study the quantum DJ
algorithm next where we show that the problem can be solved with a
single query independent of $N$. Although this exponential
speed-up is impressive when compared to the classical
deterministic approach, it is much less so when compared to the
classical probabilistic approach.

\subsection{The Quantum DJ Algorithm}
\label{subsec:quantumDJ}

The quantum DJ algorithm has been shown to solve
Problem~\ref{problem:balancedorconstant} in a single
query~\cite{DJ92}. The quantum DJ algorithm is usually studied via
its corresponding quantum circuit. We present a standard circuit
version~\cite{CEMM98} in Fig.~\ref{fig:tradQDJcircuit}.

The state represented by the lower line in
Fig.~\ref{fig:tradQDJcircuit} is referred to as the
 \emph{target} qubit. In order for easier adaption of this circuit
to the CV setting, we choose an alternative, and equivalent,
circuit formulation --- one without the target state. We take this
approach to avoid some of the difficulties the target state
introduces in \cite{BP02}. The unitary operator associated with
the oracle function changes slightly in this alternative circuit.
We discuss these differences before proceeding with analysis of
the circuit.

\begin{figure}[tbp]
            \begin{center}
            \includegraphics[width=8.5 cm]{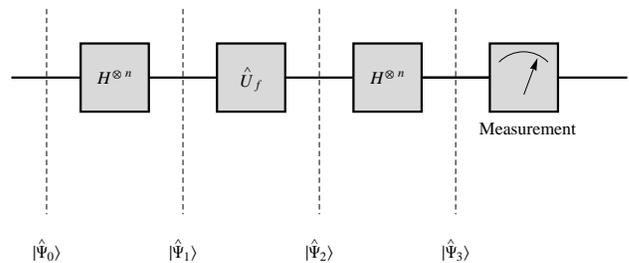}
            \end{center}
            \caption{Alternative quantum circuit implementing the discrete variable DJ Algorithm.
            Note the absence of the target qubit and the use of the operator $\hat{U}_f$ defined in
            Eq.\ref{eq:QDJTargetlessUf}}.
\label{fig:TargetlessQDJcircuit}
        \end{figure}

This simpler algorithm without the target qubit is given in
Fig.~\ref{fig:TargetlessQDJcircuit}.
Oracle application is the critical part of the algorithm. The
oracle construct originally proposed by DJ is expressed, for $x
\in \{0,1\}^n$ and $y \in \{0,1\}$, as
\begin{align}
\label{eq:QDJOracle}
    U_f:|x\rangle|y\rangle\mapsto|x\rangle|y\oplus f(x)\rangle.
\end{align}
This construction yields a matrix representation for the~$U_f$ as
a permutation matrix, hence always unitary~\cite{NC00}. With
respect to the ordered basis
\begin{align}
   B=\left\{|0\cdots0\rangle|0\rangle,|0\cdots0\rangle|1\rangle,
   \ldots,|1\cdots1\rangle|0\rangle,|1\cdots1\rangle|1\rangle\right\},\nonumber
\end{align}
the unitary matrix $U_f$ can be expressed in the following
insightful form
\begin{equation}
U_f=\left(\begin{array}{cccc}
X^{f(0\cdots 0)} &          0              &      \cdots     & 0      \\
      0            &   X^{f(0\cdots 1)}    &      \cdots     & 0      \\
     \vdots        &       \vdots            &      \ddots     & \vdots  \\
      0            &          0              &      \cdots           & X^{f(1\cdots 1)} \\
\end{array} \right),\label{eq:QDJUf}
\end{equation}
with $X$ the $2 \times 2$ $\textsf{NOT}$ operator in this case.
Here $U_f$ is a $2^{(n+1)}\times 2^{(n+1)}$ matrix, which results
from there being $2^n$ strings (the arguments of $f$) and an
additional target qubit.

The operator $U_f$ can also be expressed in the alternative ordered
basis
\begin{align}
   B'=\left\{|0\cdots0\rangle|-\rangle,|0\cdots1\rangle|-\rangle,
   \ldots,|1\cdots0\rangle|+\rangle,|1\cdots1\rangle|+\rangle\right\},\nonumber
\end{align}
as
\begin{align}
U_f=\left(\begin{array}{cc}
\hat{U}_f &          0 \\
      0            &   \openone\\
\end{array} \right),\nonumber
\end{align} with $\openone$ the $2^n \times 2^n$ identity operator.
Furthermore the operator $\hat{U}_f$ is expressed as the
$2^n\times 2^n$ matrix
\begin{equation}
\label{eq:QDJTargetlessUf}
    \hat{U}_f=\left(\begin{array}{cccc}
    (-1)^{f(0\cdots 0)} &          0              &      \cdots     & 0      \\
          0            &   (-1)^{f(0\cdots 1)}    &      \cdots     & 0      \\
         \vdots        &       \vdots            &      \ddots     & \vdots  \\
          0            &          0              &      \cdots           & (-1)^{f(1\cdots 1)} \\
    \end{array} \right),
\end{equation}
and thus provides a reduced representation for~$U_f$. It is
apparent that the operator~$\hat{U}_f$ acts on a $2^n \times 2^n$
subspace of~$U_f$ since
\begin{align}
U_f=\left(\hat{U}_f\otimes|-\rangle\langle-|
\right)\oplus\Bigl(\openone^{\otimes n
}\otimes|+\rangle\langle+|\Bigr).\nonumber
\end{align} We make the assumption that if we have the oracle $U_f$, we
we also have the oracle $\hat{U}_f$. We thus conclude that the
construction employing both control and target qubits is not
strictly necessary. That is, one could construct this algorithm
employing the $n$-qubit control state only. Apparently the choice
of representation simply depends on the nature of the actual
physical implementation.

We now present a step-by-step analysis of the alternative circuit
presented in Fig.~\ref{fig:TargetlessQDJcircuit}. We shall analyze
the CV circuit in the same steps for cross reference and
comparison.
\subsubsection{State preparation}
\label{subsubsec:stateprep}

We use the hat notation $|\hat{\Psi}\rangle$ in order to emphasize
that this analysis is of the algorithm presented in
Fig.~\ref{fig:TargetlessQDJcircuit}, which employs $n$-qubit
states and not of that presented in Fig.~\ref{fig:tradQDJcircuit},
which employs $(n+1)$-qubit states . The $n$-qubit input state of
the circuit in Fig.~\ref{fig:TargetlessQDJcircuit} is a string of
qubits prepared in $|0\cdots 0\rangle$. The next step in state
preparation is to place the state $|\hat{\Psi}_0\rangle$ into an
equal superposition of all computational basis states
\begin{equation*}
    H^{\otimes n}|\hat{\Psi}_0\rangle\mapsto|\hat{\Psi}_1\rangle
        = 2^{-n/2}\sum_{x\in\{{0,1\}^n}}|x\rangle
\end{equation*}
for~$H$ the single qubit Hadamard operator.

\subsubsection{Oracle application}
\label{subsubsec:oracleapp}

Given the definition of the reduced operator $\hat{U}_f$ defined
in Eq.~\ref{eq:QDJTargetlessUf}, its effect on the equal
superposition of basis states expressed in the state
$|\hat{\Psi}_1\rangle$ is to effectively encode the $N$-bit string
$z$ unitarily into the state $|\hat{\Psi}_2\rangle$. We express
this as
\begin{equation}
\label{eq:QDJTargetlessState2}
    \hat{U}_f|\hat{\Psi}_1\rangle\mapsto|\hat{\Psi}_2\rangle
        =2^{-n/2} \begin{pmatrix}
            (-1)^{f(0\cdots 0)}\\(-1)^{f(0\cdots 1)}\\ \vdots\\(-1)^{f(1\cdots 1)}
        \end{pmatrix},
\end{equation}
which is a convenient representation. We shall show that this
representation naturally extends to the CV setting.

\subsubsection{Measurement}
\label{subsubsec:measurement}

Measurement proceeds by first undoing the superposition created
during the state preparation step. This is achieved through the
application of the operator $\hat{U}_3=H^{\otimes n}$, which
modifies the state after oracle application
\begin{equation}
    \hat{U}_3|\hat{\Psi}_2\rangle\mapsto|\hat{\Psi}_3\rangle.
\end{equation}
The resultant state is
\begin{equation}
\label{eq:QDJState3}
    |\hat{\Psi}_3\rangle=2^{-n/2}H^{\otimes n}
        \sum_{x\in\{{0,1\}^n}}(-1)^{f(x)}|x\rangle.
\end{equation}
We rewrite Eq.~(\ref{eq:QDJState3}) with the operator $H^{\otimes n}$ expressed in terms of a
recursive definition as follows
\begin{align}
\label{eq:QDJTargetlessState3recursive}
    |\hat{\Psi}_3\rangle=&2^{-(n+1)/2}
        \begin{pmatrix}
            H^{\otimes (n-1)}   &   H^{\otimes (n-1)}       \\
            H^{\otimes (n-1)}   &   -H^{\otimes (n-1)}
        \end{pmatrix}
        \begin{pmatrix}
            (-1)^{f(0\cdots 0)}\\(-1)^{f(0\cdots 1)}\\ \vdots\\(-1)^{f(1\cdots 1)}
        \end{pmatrix}.
\end{align}
Given
\begin{equation}\label{eq:Hadamardsinglequbit}
    H=\frac{1}{\sqrt{2}}
        \begin{pmatrix}1&1\\1&-1\end{pmatrix},
\end{equation}
the combination of Eq.~(\ref{eq:QDJTargetlessState3recursive}) and
Eq.~(\ref{eq:Hadamardsinglequbit}) allows us to see that all of
the rows (and columns) of the operator $H^{\otimes n}$ have an
equal number of positive and negative ones except for the first
row, which consists entirely of plus ones. It is this feature that
permits the constant and balanced functions to be distinguished in
a single measurement.

For the two constant cases, Eq.
(\ref{eq:QDJTargetlessState3recursive}) may be expressed as
\begin{equation}
\label{eq:QDJState3Constant}
    |\hat{\Psi}_{3\text{C}}\rangle=\pm\frac{1}{2^n}
        \begin{pmatrix}1&1&\cdots &1\\
            1&-1&\cdots &-1\\
            \vdots &\vdots&\ddots&\vdots\\
            1&-1&\cdots &1
        \end{pmatrix}
        \begin{pmatrix}1\\1\\ \vdots\\1\end{pmatrix}
            =\pm\begin{pmatrix}1\\0\\ \vdots\\0\end{pmatrix}
\end{equation}
as only the first row does not result in amplitude cancellation of
the $2^n$ constant amplitude components of the state
$|\hat{\Psi}_2\rangle$. Each of the balanced functions result in
the amplitudes of the state $|\hat{\Psi}_2\rangle$ having an equal
number of positive and negative ones. This feature coupled with
action of the operator $H^{\otimes n}$ results in the first
component of the state $|\hat{\Psi}_3\rangle$ having zero
amplitude for all the balanced functions. We express this result
as
\begin{equation}
\label{eq:QDJState3Balanced}
    |\hat{\Psi}_{3\text{B}}\rangle =\pm
        \begin{pmatrix}0\\x\\ \vdots\\x\end{pmatrix},
\end{equation}
where we use the symbol~$x$ to represent that the non-zero
value(s) will land on the other $N-1$ components depending on
which of the ${N \choose N/2}$ balanced functions the oracle is
set to. It is interesting to note that the number of rows in the
state $|\hat{\Psi}_{3\text{B}}\rangle$ potentially having a
non-zero value is $N-1$ whereas the number of balanced functions
is exponential in $N$. This means that many of the balanced states
can be expressed as real-valued mixtures of the computational
basis states with the condition that the amplitude of the first
component is always zero.

For the final measurement step, we employ the projection operator~\cite{NC00}
defined  for $m\in \{0,1\}^N$ as follows
\begin{equation}
\label{eq:QDJMeasure}
    M_m=|m\rangle\langle m|.
\end{equation}
We are only concerned with the first component as discussed above,
so for the constant cases we have
\begin{equation}
\label{eq:QDJMeasureConstPrb}
    \Pr[m=(0\cdots 0)]=\left\langle\hat{\Psi}_{3\text{C}}\right|M_{(0\cdots 0)}
        \left|\hat{\Psi}_{3\text{C}}\right\rangle=1,
\end{equation}
and for all balanced cases we have
\begin{equation}
\label{eq:QDJMeasureBalPrb}
    \Pr[m=(0\cdots 0)]=\left\langle\hat{\Psi}_{3\text{B}}\right|
        M_{(0\cdots 0)}\left|\hat{\Psi}_{3\text{B}}\right\rangle=0
\end{equation}
as required.

We have completed the study of the quantum DJ algorithm in a form
that allows us to adapt readily to the CV setting. Our strategy
will be to construct a CV algorithm analogous to that shown in
Fig.~\ref{fig:TargetlessQDJcircuit} and whose operator
representation is given by
\begin{equation}\label{eq:QDJTargetlessOperator}
    |\hat{\Psi}_3\rangle=H^{\otimes n}\,\hat{U}_f \,H^{\otimes n}|0\cdots
    0\rangle.
\end{equation} This approach is simpler, and we can worry
about whether or not an implementation will require target states
when a particular implementation is considered. Before delving
into the CV algorithm, we present some background CV information.

\subsection{CV Background}
\label{subsec:CV}

The transition from DV to CV quantum information requires an
extension of Hilbert spaces to rigged Hilbert spaces~\cite{dlM05},
which allows the use of position states~$|x\rangle$ with
$x\in\mathbb{R}$ but restricts dual states to so-called `test
functions'. An inner product between position states and test
functions is meaningful, but the inner product between two
position states leads to the Dirac relation $\langle
x'|x\rangle=\delta(x-x')$, which must be treated carefully.
For~$n$ the size of Problem~\ref{problem:balancedorconstant}, the
target-less quantum DJ algorithm requires $n$ qubits, which
requires a Hilbert space of size $N=2^n$~\cite{Cav02,BCD02}. The
Hilbert space for CV problems seems quite generous in this respect
as it is infinite-dimensional.

In fact the CV Hilbert space is congruent to the space of square-integrable
complex functions over the real field $\mathcal{L}^2(\mathbb{R})$~\cite{Tan}.
A function $f:[a,b]\rightarrow \mathbb{C}$, for $[a,b]\subset\mathbb{R}$,
is in $\mathcal{L}^2(\mathbb{R})$ if
\begin{equation}
\label{eq:CV_L2}
    \int_{b}^a |f(x)|^2 \text{d}x < \infty.
\end{equation}
The inner product of two functions~$f,f'$ is
\begin{equation}
\label{eq:CV_IP}
    \langle f'|f\rangle=\int_a^b f'^*(x)f(x)\text{d}x,
\end{equation}
with positive definite norm and distance metric defined by
\begin{equation}
    ||f||=\sqrt{\langle f|f\rangle},\;
    d(f,f')=\|f-f'\|,
\end{equation}
respectively.

Typically in CV quantum information discourse, the position states~$|x\rangle$
are introduced as a basis set of the Hilbert space with each~$|x\rangle$ an
eigenstate of a position operator~$\hat{x}$, with~$x\in\mathbb{R}$.
Unfortunately the state~$|x\rangle$ does not exist in the Hilbert space;
this problem is evident in the standard inner product
\begin{equation}
\label{eq:CV_deltafunction}
    \langle x|x'\rangle=\delta(x-x').
\end{equation}
As $\delta$ is not a proper function, position states are not proper states.
Fortunately the position states are correct as a representation;
for example $f(x)=\langle x|f\rangle$ is the position representation of test function~$f$
within the context of the rigged Hilbert space.
Also Eq.~(\ref{eq:CV_deltafunction}) is meaningful in the context of distribution theory.

A rigged Hilbert space is a pair $(\mathcal{H},\Phi)$ such that
$\mathcal H$ is a Hilbert space and $\Phi$ is a vector space that
is included by a continuous mapping into $\mathcal H$:
$\Phi\subseteq\mathcal{H}$. Elements of~$\Phi$ are referred to as
`test functions', and the dual to $\Phi$ is
$\Phi^*\supseteq\mathcal H^*$, for~$\mathcal{H}^*$ dual to
$\mathcal{H}$ and~$\Phi^*$ comprising generalized functions, or
`distributions'. The inner product $\langle f'|f\rangle$ is in
$[0,1]$ for any $f'\in\Phi^*$ and for any $f\in\Phi$ \cite{dlM05}.

Note that the adaptation of the DV DJ algorithm to the CV regime
needs to be done in the context of a computational problem.  Here
the relevant problem is still
Problem~\ref{problem:balancedorconstant}, and the notion of the
oracle remains unchanged.  Thus, in the CV case, our task is still
to determine whether the function~$f_z$ belongs to the set of
constant functions or to the set of balanced functions.

\section{CV Representation of the DJ Problem}
\label{sec:CVrep}

We begin by giving a strategic overview in order to convey the key
concepts of our approach to developing a CV computation model. We
follow this with a subsection giving some preliminary definitions
allowing us to set the stage for detailed analysis. We then
proceed with a step-by-step analysis of our CV DJ algorithm.

\subsection{Strategy Overview}

Although we are now working with CV, instead of DV, quantum
information, the computational problem to be solved remains
Problem~\ref{problem:balancedorconstant}.  In other words, we want
to learn whether the function~$f_z$ is constant or balanced with
as few oracle queries as possible.  Another way to think of this
is that we wish to determine the index $b \in \{0,1\}$ such that
$f_z \in S_b$.  We now give a conceptual overview of our model for
CV quantum computation of the DJ problem, which we follow later
with a rigourous treatment.

In our model of CV quantum computation, we will use the continuous
position and momentum variables of a particle. For $x,p\in
\mathbb{R}$, we use the particle's position wave
function,~$\phi(x)$, to describe where the particle is
concentrated and the particle's momentum wave
function,~$\tilde{\phi}(p)$, to describe its momentum
distribution. The position and momentum wave functions are Fourier
transform pairs, and the relationship between the particle's
position and its momentum is governed by Heisenberg's uncertainty
principle.

\begin{figure}[tbp]
            \includegraphics[width=8.5 cm]{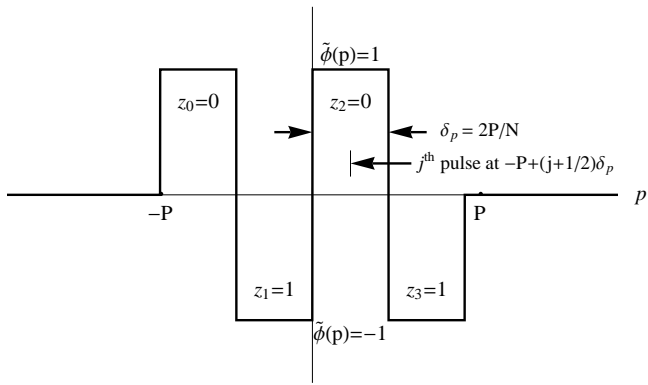}
            \caption{Illustration of the concept for encoding an~$N$-bit string
             in a region of momentum extending from~$-P$ to~$+P$ using the
             ~$N=4$,~$z=0101$ example. Note that each of the bits $z_j$ are uniquely
             represented.}
\label{fig:CVDJAlgEncodeConcept}
\end{figure}

There are many position and momentum wave function pairs on which
we could base our computational model. We select our particular
pair as follows. First, we wish to encode the unknown $N$-bit
string, $z$, in the momentum domain. We do so because encoding in
the momentum domain is the continuous analogue of the discrete
case, where encoding is performed on an equal superposition of
computational basis states. Second in order to fix one of the
degrees of freedom of the problem, we want each of the bits
comprising the string $z$ to be unambiguously represented in the
momentum space. By unambiguous we mean that each of the bits are
represented by equal-sized, non-overlapping, contiguous regions in
the momentum space.

Since we want each of the $N$-bits comprising the string to be
represented unambiguously, we naturally think of each bit as being
manifested by a finite-width square pulse whose position in
momentum space represents the bit position in the string $z$ and
whose magnitude represents the bit value. Continuing along this
line of reasoning to the representation of the entire string $z$,
we can imagine we have a region of momentum extending from $-P$ to
$+P$. All the contiguous momentum pulses within this region thus
have ``width" $\delta_p=2P/N$, and for $j\in\{0,N-1\}$, the
$j^{\textup{th}}$ momentum pulse is centred at position
$-P+(j+1/2)\delta_p$ and takes on value $(-1)^{z_j}$. We
illustrate this concept in Fig.~\ref{fig:CVDJAlgEncodeConcept} for
a particular~$N=4$ case.

\begin{figure}[tbp]
            \includegraphics[width=8.5 cm]{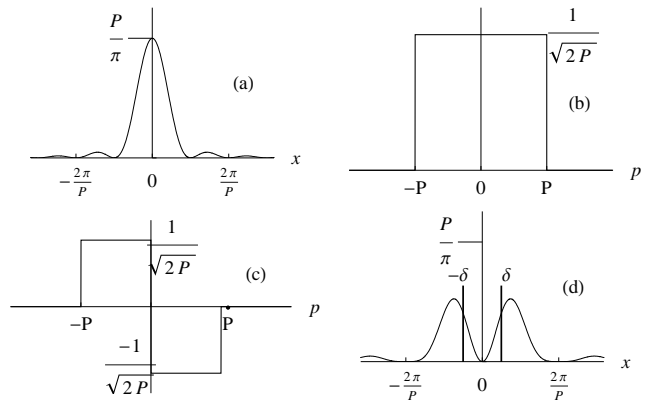}
            \caption{An illustrative overview of the four stages of our conceptual CV DJ algorithm:
             (a) The probability distribution of the input state wave function
            positioned at~$x=x_0$ is that of a sinc function.
            (b) The Fourier transform of the position state wave function is a
            ``pulse" function in the momentum domain, which acts as the encoding ``substrate".
            (c) The $N$-bit string~$z=0\cdots01\cdots1$ modulates this momentum ``substrate".
            (d) The inverse Fourier transform of the encoded ``square wave" produces a
            ``generalized" sinc function whose infinite position extent necessitates an optimal measurement
             ``window" parameterized by~$\pm \delta$.}
              \label{fig:CVDJAlgOverview}
\end{figure}The picture that thus emerges is that each of the $2^N$ possible
strings may be represented by a uniquely shaped ``square wave"
having extent $\pm P$ comprising $N$ pulses each of width $2P/N$
and having magnitude $\pm 1$. With the encoding concept clear, we
conceptually illustrate the four key stages of the algorithm in
Fig.~\ref{fig:CVDJAlgOverview}. We begin with a position wave
function centered at~$x=x_0$ and illustrated in
Fig.~\ref{fig:CVDJAlgOverview}~(a). Note that this position wave
function is a sinc function since sinc/pulse functions are Fourier
transform pairs. In Fig.~\ref{fig:CVDJAlgOverview}~(b), we present
the momentum wave function, a pulse function, which acts as the
``substrate" into which the $N$-bit strings are encoded. In
Fig.~\ref{fig:CVDJAlgOverview}~(c), the pulse function is encoded
with the particular $N$-bit string ~$z=0\cdots01\cdots1$. Finally,
the inverse Fourier transform of this ``square wave" is presented
in Fig.~\ref{fig:CVDJAlgOverview}~(d). Since the inverse Fourier
transforms of finite pulses in the momentum domain have infinite
extent in the position domain, we need to limit the extent of our
measurement to $\pm \delta$.

In summary, we see that our algorithm will need the parameters
$N$, $P$ and $\delta$. We note that as $N$ gets large, the
individual pulse width associated with a single bit becomes small
appearing to pose a limit on the maximum value of $N$. We will
return to this issue once we have determined the relationship
between $P$ and $\delta$.

There are many potential models for quantum computation in a CV
setting. We have chosen to study one where we unambiguously encode
an N-bit string into the continuous momentum variable of a
particle. Within the constraints of this model, we will show that
the CV DJ is necessarily probabilistic and prove an upper bound on
the query complexity of the CV DJ problem.

We speculate that we can't do better than this. For example if the
momentum/position pair are described by Gaussian/Gaussian
functions, as would be the case for the physically meaningful
states of quantum optics, imperfect encoding of the $N$-bit string
in the momentum domain will result in increased position error.
Whether or not this will in turn impact the ``big Oh"
representation of the query complexity requires further research
as does a general proof of a lower bound. The challenge will be to
show that another strategy can do better than the model described
herein.

\subsection{Algorithm Preliminaries}

We now proceed to formalize some of the concepts presented in the
previous subsection. Here we describe a `natural' way of encoding
a finite-dimension, $N$-bit string in a continuous domain. We
define the following function, along with its Fourier dual, to
help us achieve this end.

For $P>0$, the `top hat' function
\begin{align}
\label{eq:tophat}
    \sqcap(p;P,P_0)&=\langle p|\sqcap(P,P_0)\rangle\nonumber\\
        &=\frac{1}{\sqrt{2P}}
            \left\{
                \begin{array}{ll}1,&\text{if}
                \quad p\in[P_0-P,P_0+P]\\0,&\text{if}\quad p\notin[P_0-P,P_0+P]\end{array}
            \right.
\end{align}
will be especially useful in bridging the gap between DV and~CV
quantum information because
\begin{equation}
\label{eq:limab}
    \lim_{P\rightarrow 0}\sqcap\left(p;P,P_0\right)=\delta(p-P_0),
\end{equation}
so the state $|\sqcap\rangle$ is, in some sense, a momentum
eigenstate $|p=P_0\rangle$ in the limit $P\rightarrow 0$. The
inverse Fourier transform of the function $\text{e}^{\imath
x_0}\sqcap(p;P,P_0)$ is
\begin{align}
    \phi(x)=&\langle x | \phi\rangle \nonumber \\
        \equiv & \pi^{-1/2}\text{sinc}(P(x-x_0);P_0) \nonumber \\
        =&\frac{\text{e}^{\imath
        P_0}\sin\left(P(x-x_0)\right)}{\sqrt{P\pi}(x-x_0)},
\end{align} where $x_0$ defines the position of the sinc function.
The limit of $\phi(x)$ as $P$ goes to $\infty$ yields
$\delta(x-x_0)$. The position eigenstate $|x=x_0\rangle$ is
likewise formed in the limit $P\rightarrow \infty$.

Now imagine we want to sum a contiguous string of ``pulses"
described by the top hat function (\ref{eq:tophat}) with all
pulses having width $\delta_p$ and the $j^\textup{th}$ pulse
having complex amplitude $\psi_j$. This results in the composite
function
\begin{equation*}
\psi(p)=\sum_j\psi_j\sqcap\left(p;-P+j\delta_p,-P+(j+1)\delta_p\right).
\end{equation*}
This function can also be used as a basis \cite{Tan} of CV kets in
Dirac notation as
\begin{equation*}
    |\psi\rangle=\int_{-\infty}^\infty\text{d}p\langle
    p|\psi\rangle|p\rangle,
\end{equation*}
thus allowing us to encode quantum information in the CV domain.
Note that $\psi(p)$ is the complex amplitude for real-valued $p$.
This affords a consistent way of encoding a discrete wave function
over a continuous domain.

Before proceeding with a formal analysis of the algorithm, we give
an overview of our proof strategy. The oracle is either set to one
of two constant strings or to one of ${N\choose N/2}$ balanced
strings. A string and its complement have indistinguishable
probability distributions, so there are a total of one constant
probability distribution plus $\frac{1}{2}{N\choose N/2}$ balanced
probability distributions representing the possible oracle
settings. In order to simplify the analysis, we wish to replace
this exponential number of balanced probability distributions with
a single ``worst-case" balanced probability distribution. Thus we
seek a particular balanced string (and its complement) whose
probability distribution is most likely to ``fool" us into
concluding it is a constant string.

Intuitively, the balanced strings that have the fewest number of
changes between adjacent bits in the interval $[-P, P]$ will be
the most ``constant like" of the balanced strings. There are no
balanced strings with zero changes - this is the key feature that
separates the constant strings from the balanced strings. There is
however, a single pair of balanced strings having only one change.
These strings exhibit the feature that the first $N/2$ bits are
constant and the second $N/2$ bits are the complement of the
first. We call these strings the anti-symmetric balanced (ASB)
strings. One of these two strings is illustrated in
Fig.~\ref{fig:CVDJAlgOverview}~(c). Note that all other balanced
strings have more than one change.

Our proof strategy begins by making the assumption that the ASB
case is the ``worst case" of all balanced cases. We use this
assumption to determine the optimum value of $\delta$, which is
the extent of our measurement in the position domain and is
illustrated conceptually in Fig.~\ref{fig:CVDJAlgOverview}~(d).
Given this optimum value of $\delta$, we then prove by induction
that the worse balanced case is indeed the ASB case.

\subsection{The CV Quantum DJ Algorithm}

\begin{figure}[tbp]
            \begin{center}
            \includegraphics[width=8.5 cm]{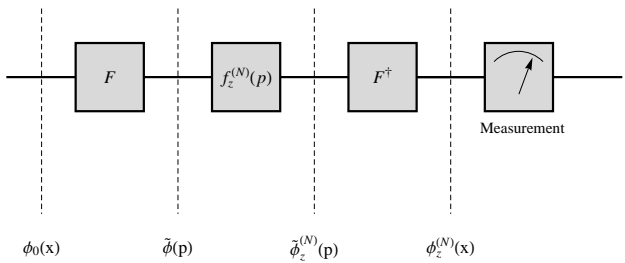}
            \end{center}
            \caption{Quantum circuit implementing the CV DJ algorithm
            without the use of the target state.}
\label{fig:TargetlessCVDJcircuit}
\end{figure}

Our strategy is to create a CV analogue of the alternative
formulation of the discrete DJ algorithm presented in
Fig.~\ref{fig:TargetlessQDJcircuit}. The CV extension of this is
presented in
Fig.~\ref{fig:TargetlessCVDJcircuit}.
Our construction of a CV DJ algorithm employs some of Braunstein
and Pati's techniques~\cite{BP02} and avoids the pitfalls. In
particular, we employ position states as a logical representation
(states in~$\Phi^*$) analogous to the discrete computational basis
states. Encoding is not, however, into the position states but
rather into test functions~$f_z\in\Phi$ with $z\in\{0,1\}^N$.
Furthermore we employ a Fourier transform to operate as a CV
version of the DV Hadamard transform (extending the Hadamard
transformation to the CV case is not unique~\cite{GKP01,BdGS02}).

For~$x$ the canonical position and $p$ the canonical momentum,
the Fourier transform maps a function $\phi(x)$ to its dual $\tilde{\phi}(p)$
according to~\cite{Bra86}
\begin{equation}
        F:\phi(x)\mapsto\tilde{\phi}(p),
\end{equation}
such that
\begin{equation}
    \tilde{\phi}(p)=\frac{1}{\sqrt{2\pi}}\int_{-\infty}^{\infty}\text{d}x\,\text{e}^{\imath p x}\phi(x)\nonumber,
\end{equation}
and
\begin{equation}
    \phi(x)=\frac{1}{\sqrt{2\pi}}\int_{-\infty}^{\infty}\text{d}p\,\text{e}^{-\imath p
    x}\tilde{\phi}(p)\nonumber
\label{eq:CVFTDefn}.
\end{equation} Note that we make use of the momentum variable $p$ as the Fourier dual
of the position variable $x$.
The function $\phi$ can be a test function in $\Phi$ and $\phi(x)$ is the inner product of
$\phi$ with the position in $\Phi^*$: $\phi(x)=\langle x|\phi\rangle$.
The momentum state $|p\rangle$ is the Fourier transform of~$|x\rangle$ and
$\tilde{\phi}(p)=\langle p|\phi\rangle$.

With these concepts in order, we now proceed through the CV DJ
algorithm analogous to the three steps in the DV DJ algorithm. The
function notation~$\phi(x)$ and $\tilde{\phi}(p)$ is more
convenient here rather than the Dirac notation in the previous
section.

\subsubsection{State preparation}
\label{subsubsec:CVstateprep}

We have argued previously that we need the Fourier transform of
the input state to be the top hat function defined in
Eq.~(\ref{eq:tophat}). We add several conditions that do not take
away from the generality of the solution. First, we want the top
hat to have zero phase, which gives $x_0=0$ and to be centred at
$P_0=0$. Second, we want the pulse to have extent $\pm P$. This
gives us the simplest form of the sinc function for the initial
state
\begin{align}
    \phi_0(x)=\frac{\sin(P x)}{\sqrt{\pi P}\, x}.
\end{align}
We note that the limit of $\phi_0(x)$ as $P\rightarrow\infty$
gives a $\delta(x)$. Thus we can think of the quantity $P$ as
playing the role of the standard deviation in a Gaussian
distribution.

The final step in state preparation is to perform the Fourier
transform, which yields the top hat function with extent $\pm P$
\begin{equation} \label{eq:puremomentum}
   \tilde{\phi}(p)
        =\frac{1}{\sqrt{2P}}
            \left\{
                \begin{array}{ll}1,&\text{if}\,\,p\in[-P,P]\\0,&\text{if}\,\,p\notin[-P,P].\end{array}
            \right.
\end{equation} This function forms the raw substrate, which will
be `modulated' by the individual $N$-bit strings $z$.
\subsubsection{Oracle application}
\label{subsubsec:CVoracleapp}

We perform encoding by partitioning the real numbers representing
momentum into non-overlapping, contiguous and equal-sized bins. In
this digital-to-analogue strategy, the width of each $p$-bin
is~$2P/N$, and
\begin{equation}
    \sqcap_i^{(N)}(p)=\left\{
            \begin{array}{ll}
                1,&\frac{p}{P}\in\left[-\left(1-2\frac{N-1-i}{N}\right),
                    -\left(1-2\frac{N-i}{N}\right)\right]\\0,&\text{otherwise.}
            \end{array}
        \right.
\end{equation}
The oracle encodes the index~$z$ into the function $f_z$ as follows:
\begin{equation}
\label{eq:fz-encoding}
    f_z^{(N)}(p)=\sum_{i=0}^{N-1} (-1)^{z_i}\sqcap_i^{(N)}(p),
\end{equation}
where the factor~$(-1)^{z_i}$ serves to modulate the phase of the top hat function
according to the bit value.

\begin{figure}[tbp]
            \includegraphics[width=8.5 cm]{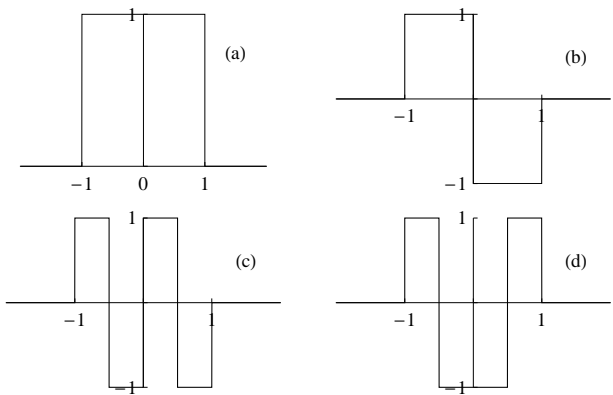}
            \caption{Encoded functions~$f^{(4)}_z(p)$ (a)~$z=0000$, (b)~$z=0011$,
                (c)~$z=0101$, and (d)~$z=0110$.}
\label{fig:CVDJSqWaveneq2}
\end{figure}

\begin{example}
Consider the case $n=2$; hence $N=2^2=4$. As one case, the
function corresponding to the four-bit string $0011$ is
\begin{equation}
\label{eq:example-0011}
    f_{0011}^{(N)}(p)
        = \sqcap_0^{(N)}(p)+\sqcap_1^{(N)}(p)
            -\sqcap_2^{(N)}(p)-\sqcap_3^{(N)}(p).
\end{equation}
The only two four-bit strings yielding constant functions would be
$0000$, for which the function is identically unity over the whole
domain~$[-P,P]$, and $1111$, for which the function is identically
$-1$ over~$[-P,P]$. Four cases are presented in
Fig.~\ref{fig:CVDJSqWaveneq2}. We refer to the function
$f_{0011}(p)$ as the ``lowest-order" antisymmetric balanced wave
as it has just one zero crossing in $[-P,P]$.

\end{example}
In the limit that $N\rightarrow\infty$ with~$P$ fixed,
$\sqcap_i(p)\mapsto\delta(p-p_i)$ for $p_i$ the midpoint of the
$i^\text{th}$ bin. The limit $N\rightarrow\infty$ thus gives a
prescription for approaching a continuous variable representation
where the $z$ index seems to approach a continuum; however this
limit yields a countable, rather than uncountable, set~$\{z\}$,
and the finite domain $[-P,P]$ has important ramifications on the
nature of the functions corresponding to Fourier transforms
of~$\sqcap_i(p)$. We express the state after encoding as
\begin{equation}
\label{eq:encodedmomentum}
    \tilde{\phi}_{z}^{(N)}(p)
        = f_{z}^{(N)}(p) \tilde{\phi}(p),
\end{equation}
where we observe the ``modulating'' effect of the encoded string
$f_{z}$ on the momentum ``substrate'' $\tilde{\phi}(p)$.

In the context of the digital-to-analogue strategy, the constant
functions are analogous to direct current (DC) signals and the
balanced functions to alternating current (AC) signals. The number
of zero-crossings corresponds to frequency information, and the
question of whether the output is balanced or constant is
essentially a problem of querying whether there is a non-zero
frequency component of the output signal. As noted previously, the
ASB function has the lowest frequency component. We now proceed to
analyze the measurement stage.
\subsubsection{Measurement}
\label{subsubsec:CVmeasurement}

We have the strings $z\in\{0,1\}^N$ encoded into the momentum
state~(\ref{eq:encodedmomentum}). The next step prior to the final
measurement is to take the inverse Fourier transform of this pulse
train. For $z_j$ the $j^\textup{th}$ bit of $z$, this is expressed
as
\begin{align}
\phi_{z}^{(N)}(x)=&F^\dag\left(\tilde{\phi}_{z}^{(N)}(p)\right) \nonumber\\
=&\frac{\imath}{2\sqrt{P\pi
}\,x}\sum_{j=1}^{N}(-1)^{z_j}\nonumber\\
&\times\left(\text{e}^{\imath\left(\frac{N-2j}{N}\right) P
x}-\text{e}^{\imath\left(\frac{N-2(j-1)}{N}\right)P
x}\right).\label{eq:CVDJPhaser1}
\end{align}
The expression given in Eq.~(\ref{eq:CVDJPhaser1}) can be
simplified to yield
\begin{align}
\phi_{z}^{(N)}(x)&=\frac{\sin(P x/N)}{\sqrt{P \pi
}\,x}\sum_{j=1}^{N}(-1)^{z_j}\left(\text{e}^{\imath\left(\frac{N-(2j-1)}{N}\right)P x}\right)\nonumber\\
&=\frac{\sin(P x/N)}{\sqrt{P\pi
}\,x}\sum_{j=1}^{N}(-1)^{z_j}\text{e}^{\imath\varphi_j(x)},\label{eq:CVDJPhaser2}
\end{align}where we have defined
\begin{equation}\label{eq:varphidef}
\varphi_j(x)=\left(\frac{N-(2j-1)}{N}\right)Px.
\end{equation}
We see that the magnitude of an individual generalized sinc
function, $\phi_{z}^{(N)}(x)$, is determined by a vector sum of
$N$ phasors, which is  modulated by a particular $N$-bit string
$z$.

Note that the phasors, $\text{e}^{\imath\varphi_j(x)}$, are
equiangular divisions of the angular interval
\begin{equation}
[+(N-1)Px/N,-(N-1)Px/N]\nonumber,
\end{equation} and they exhibit the pairwise
complex conjugate property $\varphi_j(x)=-\varphi_{N+1-j}(x)$.
\begin{figure}[tbp]
            \includegraphics[width=7 cm]{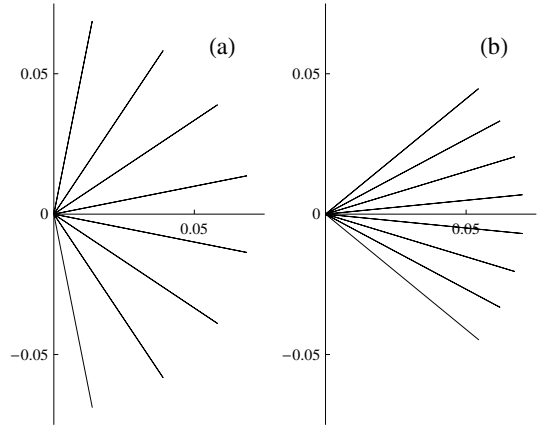}
            \caption{The phasors $\text{e}^{\imath\varphi_j x}$ defined in
            Eq.~(\ref{eq:CVDJPhaser2})
             for $N=8$ (a) $x=\pi/2$ phasors range between $\pm 7\pi/16$ in steps of $\pi/8$,
             and(b)
$x=\pi/4$ phasors range between $\pm 7\pi/32$ in steps of
$\pi/16$.} \label{fig:CVDJSincPhasers}
 \end{figure} In Fig.~\ref{fig:CVDJSincPhasers}, we present the phasors for
$N=8$ with $x=\pi/2$ and $x=\pi/4$ to illustrate these features.
Note that the phasors are added constructively or
de-constructively depending on the phase of the angles, which
results from the term $(-1)^{z_j}$. This effect defines the
magnitude of the resulting sinc function.

We note that the only functions with $\phi^{(N)}_z(0)\neq 0$ are
the two constant sinc functions. This is clear given that $
\sum_{j=1}^{N}(-1)^{z_j}=\pm N$ for the two constant cases, and $
\sum_{j=1}^{N}(-1)^{z_j}=0$ since for all balanced cases, the
latter sum always resolves to $N/2-N/2=0$. This feature of the set
of ${N\choose N/2 }+ 2$ sinc functions represented by
$\phi^{(N)}_z(x)$ defined by Eq.~(\ref{eq:CVDJPhaser2}) implies
the strategy for measurement that will distinguish between the
constant and balanced cases.

In order to refine this strategy, we focus on two cases. The first
of these cases is for the two constant functions for which
Eq.~(\ref{eq:CVDJPhaser2}) gives the probability distribution
\begin{align}
\mathcal{P}_{\text{\tiny{C}}}(x)=|\phi_{\text{\tiny{C}}}^{(N)}(x)|^2=\frac{\sin
^2(P x)}{P \pi x^2} \label{eq:CVDJState3PDFConst},
\end{align} where we have
 \begin{equation}
 \text{\small{C}}\in\left\{\underbrace{0\cdots 0}_{N},
\underbrace{1\cdots 1}_{N}\right\}.\nonumber
\end{equation}
The second case deals with the two balanced functions having the
lowest `frequency' content, which occurs when the first $N/2$ bits
and the last $N/2$ bits have opposite values.

We think of these two balanced functions as having the lowest
frequency content since Eq.~(\ref{eq:fz-encoding}) has a single
zero crossing in the interval $[-P,P]$ for these two balanced
strings only. All other balanced strings have more than one zero
crossing and thus higher frequency content. For this pair of
balanced functions, which we call the antisymmetric balanced (ASB)
functions, Eq.~(\ref{eq:CVDJPhaser2}) gives the probability
distribution
\begin{align}
\mathcal{P}_{\text{\tiny{ASB}}}(x)=|\phi_{\text{\tiny{ASB}}}^{(N)}(x)|^2=\frac{(\cos
(P x)-1)^2}{P \pi x^2} \label{eq:CVDJState3PDFABal},
\end{align} where we have
\begin{equation}\text{\small{ASB}}\in\left\{\underbrace{0\cdots
0}_{N/2}\underbrace{1\cdots 1}_{N/2}, \underbrace{1\cdots
1}_{N/2}\underbrace{0\cdots 0}_{N/2}\right\}.
 \end{equation} Note
that of the $N \choose N/2$ balanced functions, there are many
that are also antisymmetric about the midpoint. However, we
reserve the term ASB for these two \emph{lowest-order}
antisymmetric balanced functions.

\begin{figure}[tbp]
            \includegraphics[width=8.5 cm]{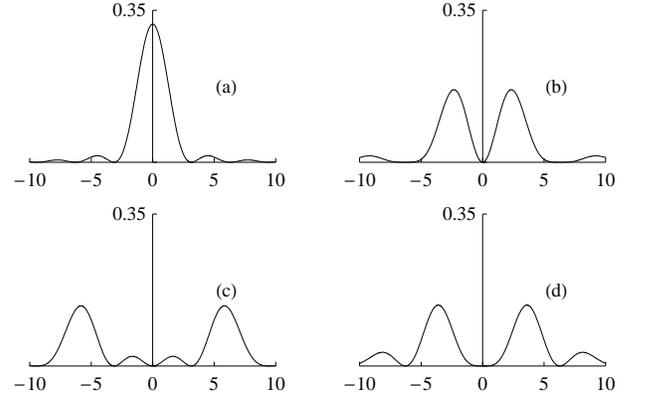}
            \caption{The probability distributions~$|\phi_{z}^{(4)}(x)|^2$ for
            (a)~$z=0000$, (b)~$z=0011$,
            (c)~$z=0101$, and (d)~$z=0110$. We clearly see that
            of the three balanced cases (b) through (d), the ASB function (b)
            has probability peaks closest to $x_0=0$.}
\label{fig:CVDJSincFtnsneq2}
 \end{figure}

We will use these cases to bound the success probability of
distinguishing between the constant and all balanced cases. We
first illustrate the concept of frequency in the following
example.
\begin{example}
Again consider the case $P=1$, $n=2$; hence $N=2^2=4$. As one
case, the function corresponding to the four-bit string $0011$ is
\begin{align}
\label{eq:example-0011-PDF}
    \phi_{0011}^{(4)}(x)
        &= \frac{\sin(x/4)}{\sqrt{\pi}x}\left(\text{e}^{\imath\frac{3x}{4}}
        +\text{e}^{\imath\frac{x}{4}}-\text{e}^{-\imath\frac{x}{4}}-\text{e}^{-\imath\frac{3x}{4}}\right)\nonumber\\
        &=\frac{-\imath\left(\cos x-1\right)}{\sqrt{\pi}x}
\end{align}
This function corresponds to the $N=4$ ASB function. The
probability distributions for the four distinct $N=4$ cases are
presented in Fig.~\ref{fig:CVDJSincFtnsneq2}. We clearly see that
of the three balanced cases, the $N=4$ ASB function has
probability peaks closest to $x_0=0$.
\end{example}

\begin{figure}[tbp]
            \begin{center}
            \includegraphics[width=6 cm]{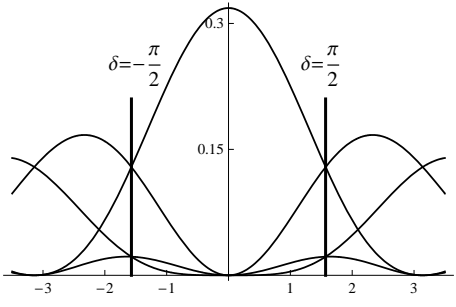}
            \end{center}
            \caption{For $P=1 $, the optimal value of $\delta=\frac{\pi}{2}$. This graph shows
            that only the Constant and the Antisymmetric Balanced Functions
            significantly contribute to probability between $\pm\delta$. }
\label{fig:CVDJoptimalmeasurelimits}
\end{figure}

 Our measurement strategy is to measure the probability
distribution in a small band around the position $x_0=0$
parameterized by $\pm \delta$. The CV analog of the projection
operator given in Eq.(\ref{eq:QDJMeasure})is defined as
\begin{align}
E_a^b=\int_{-\infty}^{\infty}D_a^b(x)|x\rangle\langle x|
\text{d}x\label{eq:CVDJMeasure},
\end{align} where
\begin{equation}
    D_a^b(x)
        =\left\{\begin{array}{ll}
        1,&\text{if}\,\,a\leq x\leq b \\
        0,& \text{otherwise.}\\
\end{array}
\right.
\end{equation}\label{eq:CVDJDoorway}Due to the symmetry of the sinc
functions about $x_0$, we set $a=-\delta$ and $b=+\delta$. We now
need to determine the optimal value of $\delta$ that will maximize
our ability to distinguish between the constant and balanced
cases. We will determine the optimum value of $\delta$ by first
assuming that the probability distribution
$\mathcal{P}_{\text{ASB}}(x)$ given by
Eq.~(\ref{eq:CVDJState3PDFABal}) dominates all other balanced
probability distributions in the region $[-\delta,\delta]$. After
using this assumption to determine a value for the optimal delta,
we will state and prove a theorem justifying our assumption. As an
illustration that our assumption is true for the $N=4$ case, we
plot the four distinct cases in
Fig.~\ref{fig:CVDJoptimalmeasurelimits}.

The ability to effectively distinguish between two random events
is proportional to the separation of the individual probabilities
of occurrence. Thus we need to select $\delta$ such that we get as
much separation between the constant distribution and the ASB
distribution as possible. Given this concept we can think that
when we make a measurement we are distinguishing between two
events, the probabilities for which we define as follows
\begin{equation}
 \text{Pr}_{\text{\tiny{Const}}}(\delta)=\text{Pr}\left[\,\left|\phi^{(N)}_z\right|^2=\mathcal{P}_{\text{\tiny{C}}}(x)\right]=
 E^{\delta}_{-\delta}\left(\mathcal{P}_{\text{\tiny{C}}}(x)\right)
 \label{eq:CVDJprobconst},
\end{equation} and
\begin{equation}
 \text{Pr}_{\text{\tiny{ASB}}}(\delta)=\text{Pr}\left[\,\left|\phi^{(N)}_z\right|^2=\mathcal{P}_{\text{\tiny{ASB}}}(x)\right]=
 E^{\delta}_{-\delta}\left(\mathcal{P}_{\text{\tiny{ASB}}}(x)\right)
 \label{eq:CVDJprobASB}.
\end{equation}

We can determine the optimum value of $\delta$ by maximizing the
expression
$|\text{Pr}_{\text{\tiny{Const}}}(\delta)-\text{Pr}_{\text{\tiny{ASB}}}(\delta)|$.
It suffices to find the value of $\delta$ for which
$\frac{\text{d}}{\text{d}\delta}|\text{Pr}_{\text{\tiny{Const}}}(\delta)-\text{Pr}_{\text{\tiny{ASB}}}(\delta)|=0$,
which may be expressed as
\begin{align}
&\frac{\text{d}}{\text{d}\delta}\left|\text{Pr}_{\text{\tiny{Const}}}(\delta)-\text{Pr}_{\text{\tiny{ASB}}}(\delta)\right|\nonumber\\
 &=\frac{\text{d}}{\text{d}\delta}\left|\int_{-\delta}^{\delta}\left(\frac{\sin ^2(P x)}{P \pi
x^2}-\frac{(\cos (P x)-1)^2}{P \pi x^2}\text{d}x\right)\right|\nonumber\\
&=\frac{\sin ^2(P \delta)}{P \pi \delta^2}-\frac{(\cos (P
\delta)-1)^2}{P \pi \delta^2}=0.\label{eq:CVDJprobabilitybound}
\end{align}This occurs where $\cos(P\delta)=\cos(P\delta)^2$ for $\delta\neq 0$,
which gives a global maximum at $\delta=\frac{\pi}{2P}$. It is
interesting to think of this result as an uncertainty relationship
\begin{align}
 P\delta=\frac{\pi}{2}.\label{eq:CVDJuncertainty}
\end{align}We shall return to this concept in our discussion in
the conclusion. We have determined the optimum value for $\delta$
based on our assumption that for $-\delta\leq x\leq\delta$ the
balanced probability distribution
$\mathcal{P}_{\text{\tiny{ASB}}}(x)$ dominates all other balanced
probability distributions. We now proceed to prove this
assumption.

In order to proceed with the proof, we define a set $\Phi$ of $m$
pairwise conjugate angles with $2m=N$. Note that $N$ is not
restricted to being equal to $2^n$ for the purpose of this proof.
Also for the purpose of this proof, we set $P=1$ and incorporate
$x$ into the definition of
$\varphi_j=\left(\frac{N-(2j-1)}{N}\right)x$ for $-\pi/2\leq
x\leq\pi/2$. We let
$\Phi=\{\varphi_1,\varphi_2,\ldots,\varphi_m,\varphi_{m+1},\ldots,\varphi_{2m}\}$
with $j=1,\ldots, m$ and note the pairwise conjugate property
$\varphi_j=-\varphi_{2m+1-j}$. Now consider
$\mathbf{S}=\sum_{j=0}^{2m}g(j)\text{e}^{\imath \varphi_j}$ where
$g:[2m]\mapsto\pm 1$ subject to the balanced condition $\sum_j
g(j)=0$, then

\begin{theorem}
Max $|\mathbf{S}|$ occurs under the specific balanced conditions
\begin{align}
&g(j)=\left\{\begin{array}{cl}
1 & \mbox{if $ 1 \leq j \leq m$} \\
-1 & \mbox{if $ m+1 \leq j \leq 2m$}, \\
\end{array}
\right.\end{align} and
\begin{align}&g(j)=\left\{\begin{array}{cl}
-1 & \mbox{if $ 1 \leq j \leq m$} \\
1 & \mbox{if $ m+1 \leq j \leq 2m$}, \\
\end{array}
\right.
\end{align} which we refer to as the asymmetric balanced
functions (ASB).
\end{theorem}

\begin{proof}
Proof is done by induction on $m$. We begin with the base case
$m=1$, $N=2$. This case is trivial since the only balanced cases
are the two ASB cases represented by the strings $\{01,10\}$. We
proceed with the base case for $m=2$, $N=4$. This case is a little
more involved. We begin by labelling the angles and phasors as
shown in Fig.~\ref{fig:CVDJoangledefintionsforproof}.

\begin{figure}[tbp]
            \begin{center}
            \includegraphics[width=6 cm]{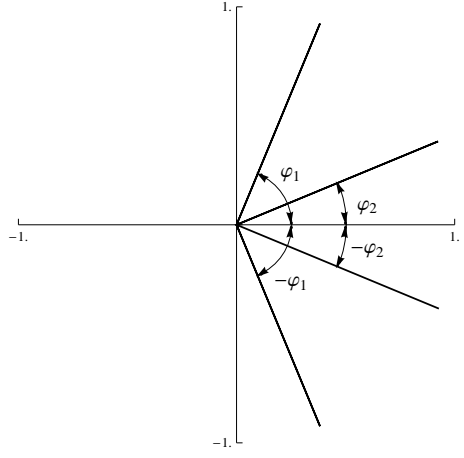}
            \end{center}
            \caption{Definition of the phasor angles for the $N=4$ base base. Note that the effect of
            varying $x$ over $[-\pi/2,\pi/2]$ simply focuses or expands the double
            angles $2\varphi_1$ and $2\varphi_2$ proportionally. }
\label{fig:CVDJoangledefintionsforproof}
\end{figure}There are ${4\choose2}=6$ balanced cases, and we have to consider
the strings $\{0011,0101,0110,1100,1010,1001\}$. Since the latter
three are complements of the first three, we have to consider only
three vector sums.

With reference to Fig.~\ref{fig:CVDJoangledefintionsforproof}, we
have
$\mathbf{S}_{\{0011\}}=\text{e}^{\imath\varphi_1}+\text{e}^{\imath\varphi_2}-\text{e}^{-\imath\varphi_1}-\text{e}^{-\imath\varphi_2}$.
We simplify and express the resultant along with the three other
cases as
\begin{enumerate}
\item$\mathbf{S}_1=\mathbf{S}_{\{\{0011\},\{1100\}\}}=\pm 2\imath(\sin(\varphi_1)+\sin(\varphi_2))$
\item  $\mathbf{S}_2=\mathbf{S}_{\{\{0101\},\{1010\}\}}=\pm 2(\cos(\varphi_1)-\cos(\varphi_2))$
\item  $\mathbf{S}_3=\mathbf{S}_{\{\{0110\},\{1001\}\}}=\pm
2\imath(\sin(\varphi_1)-\sin(\varphi_2))$.
\end{enumerate} Clearly $|\mathbf{S}_1|>|\mathbf{S}_3|$. We use
the trigonometric identities,\begin{align}
|\mathbf{S}_1|&=2\sin\left(\frac{\varphi_1+\psi_2}{2}\right)\cos\left(\frac{\varphi_1-\varphi_2}{2}\right)\nonumber\\
|\mathbf{S}_2|&=2\sin\left(\frac{\varphi_1+\varphi_2}{2}\right)\sin\left(\frac{\varphi_1-\varphi_2}{2}\right),
\end{align} to establish the relationship
between $|\mathbf{S}_1|$ and $|\mathbf{S}_2|$. We note that
$\max\left(\frac{\varphi_1-\varphi_2}{2}\right)=\left(\frac{1}{2}-\frac{1}{N}\right)x
$ and
$\min\left(\frac{\varphi_1-\varphi_2}{2}\right)=\left(\frac{1}{N}\right)x
$ for all $N$ and the specified range of $x$. Since $\cos x > \sin
x $ for $0\leq x < 1/2$, we conclude that
\begin{equation}\cos\left(\frac{\varphi_1-\varphi_2}{2}\right)>
\sin\left(\frac{\varphi_1-\varphi_2}{2}\right)\nonumber,\end{equation}
and thus $|\mathbf{S}_1|>|\mathbf{S}_2|$ for $0\leq x \leq \pi/2$.
This proves that the theorem is true for the $m=2$, $N=4$ base
case. We are now ready to prove the inductive step.

We consider two cases. Case (i) assumes every pair is balanced. By
this we mean that $g(j)=-g(2m+1-j)$. By inspection, this gives the
same result as $|\mathbf{S}_1|$ and $|\mathbf{S}_3|$ for the
$m=2$, $N=4$ case. Case(ii) assumes that Case(i) is not true and
is proved by induction. Since Case(i) is not true, there must
exist two non-balanced pairs for which $g(j)=g(2m+1-j)=+1$ and
$g(k)=g(2m+1-k)=-1$. As an illustration in the $m=4$, $N=8$ case,
the balanced string $\{01000111\}$ has this property. The
inductive step is
\begin{align}
\left|\sum_{j=1}^{2m}g(j)\text{e}^{\imath\phi_j}\right|&\leq
|\mathbf{S}\left(\{l,2m+1-l,k,2m+1-k\}\right)|\nonumber\\
&+|\mathbf{S}(w)|,
\end{align}
where $\mathbf{S}(w)$ is maximized for the $m=2$, $N=4$ base case.
Only when $|\mathbf{S}\left(\{l,2m+1-l,k,2m+1-k\}\right)|$ itself
is maximized is equality achieved and the total sum maximized.
This occurs for the ASB strings.
\end{proof}

We have established that we can bound the probabilities of
determining whether an unknown function is balanced or constant in
a single query in a CV setting. In the next section will determine
an upper bound for the query complexity of a CV algorithm in terms
of success probability in terms of the number of queries.

\section{Bounding the Query Complexity of the Continuous Variable DJ algorithm}
\label{sec:bounding} Before we bound the query complexity, we make
some important observations regarding the comparison between the
discrete DJ algorithm and the CV DJ algorithm.

First, we note that probability distributions
$\mathcal{P}_{\text{\tiny{C}}}(x)$ and
$\mathcal{P}_{\text{\tiny{ASB}}}(x)$ defined by
Eqs.~(\ref{eq:CVDJState3PDFConst})
and~(\ref{eq:CVDJState3PDFABal}) respectively, are in
$\mathcal{H}_2$, the Hilbert space of $\mathcal{L}^2(\mathbb{R})$
functions over the interval $[-\infty,\infty]$. This implies that
since we are measuring over a finite interval, the CV DJ algorithm
is necessarily probabilistic. Furthermore, we noted that $P$ and
$\delta$ are related by the uncertainty relation given in
Eq.~(\ref{eq:CVDJuncertainty}). This leads to the conclusion that
even in the limit of the improper delta function $\delta(x-x_0)$,
the CV DJ algorithm remains probabilistic. This conclusion is
contrary to that made in~\cite{BP03}.

Second, we compare the operator descriptions of the DV DJ and the
CV DJ, which we express as
\begin{align}
&|\hat{\Psi}_3\rangle=H^{\otimes n}\,\hat{U}_f \,H^{\otimes
n}|0\cdots 0\rangle\nonumber\\
&\phi_z^{(N)}(x)=F^{\dag}_p\,f^{(N)}_z(p) \,F_x\,
\phi_0(x).\label{eq:QDJandCVDJOperators}
\end{align}
The first equation represents the quantum DJ algorithm operator
expression given in Eq.~(\ref{eq:QDJTargetlessOperator}). The
second equation is the analogous CV operator expression determined
by concatenating the steps of the previous section. There is a
high degree of similarity between these two expressions, but there
are mathematical subtleties.
\begin{enumerate}
\item The CV position state $\phi_0(x)$ is not a perfect analog to
the computational basis state $|0\cdots 0\rangle$ except in the
limit. However, this limit creates a state that is not in the RHS
we argued is necessary for consistency~\cite{dlM05}.
\item The continuous
Fourier transform is not equal to the CV extension of the Hadamard
operator in a CV parameterized system with a finite Hilbert space.
It is however, a convenient extension when the Hilbert space is
infinite.
\item Finally, the
diagonal operator $\hat{U}_f$ given by Eq.
(\ref{eq:QDJTargetlessUf}) has each entry taking on the value $\pm
1$ dependent on the value of $f_{z}$. The CV analogue to this
operator is the function $f^{(N)}_z(p)$, where each of the $N$
partitions of the real interval $[-P,P]$ similarly take on the
value $\pm 1$.
\end{enumerate}

We now determine numerical values of the probabilities determined
in Eqs.~(\ref{eq:CVDJprobconst}) and~(\ref{eq:CVDJprobASB}). We
can readily calculate the probability of detecting if the function
is constant
\begin{align}
\label{eq:CVDJMeasureConstPrb}
    \text{Pr}_{\mbox{\tiny{Const}}}&=\int_{-\delta}^{\delta}\frac{\sin ^2(P x)}
            {P \pi x^2}\text{d}x \nonumber\\
        &=\frac{\cos (2 \delta P)+2 \delta P \text{Si}(2 \delta P)-1}{\delta P \pi},
\end{align}
where the sine integral is given by
\begin{equation}
    \text{Si}(z)=\int_0^z \frac{\sin t}{t}\text{d}t.
\end{equation}
Note this probability depends only on the product $P\delta$. If
the function is the lowest-order antisymmetric balanced (ASB)
\begin{align}
 \text{Pr}_{\mbox{\tiny{ASB}}}&=\int_{-\delta}^{\delta}\frac{(\cos (Px)-1)^2}{P \pi x^2}\text{d}x \nonumber\\
 &=\frac{-8 \sin
^4\left(\frac{\delta P}{2}\right)+ 4 \delta P \text{Si}(\delta
P)-2\delta P \text{Si}(2\delta P)}{\delta P \pi }.
\end{align}\label{eq:CVDJMeasureABalPrb}For $P \delta=\pi/2$, the numerical values of these two
probabilities are
\begin{equation}
\text{Pr}_{\mbox{\tiny{Const}}}=\frac{2\left(\pi\,\text{Si}\,(\pi)-2\right)}{\pi^2}\approx
0.77 , \label{eq:CVDJprobfigureconst}\end{equation} and
\begin{equation}
\text{Pr}_{\mbox{\tiny{ASB}}}=\frac{4\pi\,\text{Si}\,(\pi/2)-2\pi\,\text{Si}\,(\pi)-4}{\pi^2}\approx
0.16.\label{eq:CVDJprobfigureasb}
\end{equation}
Given this probabilistic nature of the CV DJ algorithm, we need to
develop a strategy to bound the error probability. We will employ
the technique sometimes called \emph{probability amplification}
\cite{Ad04, Can01}.

Our strategy will be to make $m$ repetitions of the CV DJ
algorithm where we assume that the oracle is set to the same
function for each of the repetitions. Each repetition ends with a
measurement. From this sequence of measurements we want to
determine whether the unknown function is balanced or constant
with high probability.
\begin{theorem} An error of $O(\text{e}^{-m})$ can be achieved by making
$O(m)$ repetitions of the CV DJ algorithm.
\end{theorem}
\begin{proof}
We will adopt the convention that when we make a query to the CV
DJ algorithm we either detect something (algorithm returns a 1),
or we do not (algorithm returns a 0). We can thus treat multiple
queries as a sequence of Bernoulli trials \cite{CLR90}. We assume
that we have set our measurement limits to the optimal
$\pm\delta$. The two events we are trying to uncover are the
constant cases where, for ease of calculation we set the
probability of detecting something is
$\text{Pr}_{\mbox{\tiny{C}}}\geq 3/4$, and the balanced cases
where the probability detecting something is
$\text{Pr}_{\mbox{\tiny{B}}}\leq1/4$. Note that we have set the
probabilities to these rational numbers for illustrative purposes
and to simplify the calculation. We can make this arbitrary
setting, and we will get the same result as long as the
probabilities are bound from $1/2$ by a constant.

If each measurement is based on an independent preparation of the
state $\phi_0(x)$, then each of the queries are independent. After
a series of $m$ queries, we can use the Chernoff bounds of the
binomial distribution to amplify the success probability
~\cite{CLR90, Can01}. The simplest (but somewhat weak) Chernoff
bound on the lower tail is given by~\cite{Can01} as
\begin{align}
 \text{Pr}[X<(1-\epsilon)\mu)]<\text{e}^{-\frac{\mu \epsilon^2}{2}},\label{eq:CVDJChernoffweakL}
\end{align}
and on the upper tail as
\begin{align}
 \text{Pr}[X>(1+\epsilon)\mu)]<\text{e}^{-\frac{\mu \epsilon^2}{4}},\label{eq:CVDJChernoffweakU}
\end{align}
where $\mu$ is the expected mean of the resulting binomial
distributions after $m$ queries, and $\epsilon$ is the relative
distance from the respective means.

First, we bound the lower tail corresponding to the distribution
of the constant case for which we have
$\mu=m\,\mathcal{P}_{\mbox{\tiny{C}}}$. Here we set
$\epsilon=\frac{1}{3}$, which expresses the probability for the
value being less than half way between the two means as
$\text{Pr}[X<(m/2)]<\text{e}^{-\frac{m}{24}}$. Similarly, we bound
the upper tail for the balanced case for which we have
$\mu=m\,\mathcal{P}_{\mbox{\tiny{B}}}$. Here we set $\epsilon=1$,
which expresses the probability for the value being greater than
half way between the two means as
$\text{Pr}[X>(m/2)]<\text{e}^{-\frac{m}{16}}$. Clearly the success
is worse for the lower tail allowing us to bound the success
probability of the CV DJ algorithm after $m$ queries as
\begin{align}
\text{Pr}[\text{\text{Success}}]\geq1-\text{e}^{-\frac{m}{24}}.\label{eq:CVDJSuccessProb}
\end{align}
This gives an error probability that is $O\left(\text{e}^{- m
}\right)$ as required.
\end{proof}
We note that this is of the same order as the exponentially good
success probability we have for the classical probabilistic
approach given by Eq.~(\ref{eq:ClassDJRan}). Also note that this
query complexity is independent of the value of $N$. We have made
no attempt to obtain a tighter bound preferring to show only that
the success probability of the CV DJ algorithm is of the same
order of that of the classical probabilistic approach to solving
the DJ problem.
\section{Conclusions}

In this paper we have presented a rigorous framework for the
analysis of the DJ oracle identification problem in a CV setting.
The rigged Hilbert space (RHS) affords a consistent transition
from the traditional discrete Hilbert space to the CV setting. Our
framework allows us to define a consistent way of encoding $N$-bit
strings into functions over the real numbers.

We have used this framework, and the selection of the sinc/pulse
Fourier transform pair, to prove that a CV implementation of the
DJ algorithm cannot provide the exponential speed-up of its
discrete quantum counterpart. Additionally, we have presented a
bounded-error, upper bound on the query complexity of the DJ
problem within the constraints of our model. The lack of speed-up
results from an uncertainty principle between the ability to
encode perfectly in a continuous representation and the subsequent
inability to measure perfectly in the Fourier-dual representation.
This uncertainty relationship is manifest in
Eq.~(\ref{eq:CVDJuncertainty}), which relates $P$, the encoding
extent, to $\delta$, the measurement extent. A natural extension
of this work would be prove a lower bound perhaps exploring the
techniques along the lines of \cite{HS05} from the perspective of
different Fourier transform pairs.

This uncertainty relationship appears to be a natural feature of
the CV setting, but it could also be used to advantage. There is
likely to be oracle function symmetries that are particularly
suited to different CV settings. For example in
Sec.~\ref{sec:CVrep}, we showed that balanced functions with a
higher number of zero crossings create sinc functions with
frequency components further away from $x_0$. It appears that an
oracle identification problem designed to separate balanced
functions according to frequency separation could be implemented
in a CV setting and possibly provide advantage over classical or
discrete quantum settings.

Furthermore, it would be interesting to classify the balanced
functions from the perspective of different coherent
states~\cite{Per86} in CV parameterized settings of both finite
and infinite dimensions. The former would naturally involve the
study of the coherent spin systems~\cite{ACGT72}. Furthermore the
use of squeezed spin states should be studied~\cite{KU93}.
Infinite dimension systems would naturally involve the study of
implementations involving the coherent states of quantum
optics~\cite{ACGT72,Le97}.

Additionally, we have set up this framework in a manner that
should allow any oracle identification problem to be analyzed in a
similar manner in the CV setting. An implementation of a discrete
quantum oracle, for example~\cite{BV97,CvDNT99}, requires a
unitary operator representing the oracle. Provided we can create a
diagonal representation of this oracle along the lines of
$\hat{U}_f$ given in Eq.~(\ref{eq:QDJTargetlessOperator}), our
framework will naturally extend to it. Of course we need to be
able to create an implementation of these oracles and that remains
an important open question.

Other avenues of the extension of this framework include CV
implementations of other hidden subgroup problems. The solution of
Simon's problem~\cite{Sim97} in this setting would be an obvious
starting point as would the exploration of a CV implementation of
Shor's algorithm~\cite{Sho97}. Additionally, the CV framework
could be extended to include analysis of noisy oracles along the
lines of~\cite{AIKRY04, AC02}.

In closing we note that the transition from a discrete quantum
information setting to a CV setting has many subtleties. In
particular the improper delta functions must not be used. Limiting
behaviour can be explored, but only if the limits are taken from
the perspective of functions defined in the rigged Hilbert space.

\section*{Acknowledgements}
We appreciate financial support from the Alberta Ingenuity Fund
(AIF), Alberta's Informatics Circle of Research Excellence
(\emph{i}CORE), Canada's Natural Sciences and Engineering Research
Council (NSERC), the Canadian Network Centres of Excellence for
Mathematics of Information Technology and Complex Systems
(MITACS), and General Dynamics Canada. PH is a Scholar and BCS is
an Associate of the Canadian Institute for Advanced Research
(CIFAR).


\begin{thebibliography}{10}

\bibitem {NC00}
    M. A. Nielsen and I. L. Chuang,
    \emph{Quantum Computation and Quantum Information}
    (Cambridge University Press, Cambridge, UK, 2000).

\bibitem {GKP01}
    D. Gottesman, A. Kitaev, and J. Preskill, ``Encoding a qubit in an oscillator '',
    \pra \textbf{64}, 012310 (2001).

\bibitem {BP03}
    S. L. Braunstein and A. K. Pati, eds.
    \emph{Quantum Information with Continuous Variables},
     (Kluwer, Dordrecht, 2003); arXiv: quant-ph/0207108.

\bibitem {FSBFK98}
    A. Furusawa, J. L. S\o rensen, S. L. Braunstein, C. A. Fuchs, H. J. Kimble, and E. S.
    Polzik, ``Unconditional quantum teleportation'', Science \textbf{23}, Vol. 282. no. 5389, pp. 706-709,
    (1998)

\bibitem {GG02}
    F. Grosshans and P. Grangier, ``Continuous variable quantum cryptography using coherent states'',
    \prl \textbf{88}, 057902-1-057902-4 (2002).

\bibitem {AFKLL08}
    J. Appel, E. Figueroa, D. Korystov, M. Lobino, and A. I. Lvovsky,
    ``Quantum memory for squeezed light'', \prl \textbf{100}, 093602 (2008).

\bibitem {AYANTFK07}
    D. Akamatsu, Y. Yokoi, M. Arikawa, S. Nagatsuka, T. Tanimura, A. Furusawa, and M. Kozuma,
     ``Ultraslow propagation of squeezed vacuum pulses with electromagnetically induced transparency'',
      \prl \textbf{99}, 153602 (2007).

\bibitem {Br98}
    S. L. Braunstein, ``Error correction for continuous quantum variables'', \prl \textbf{80}, 4084 (1998).

\bibitem {EP02}
    J. Eisert and M. B. Plenio, ``Distilling Gaussian states with Gaussian operations is impossible'',
    \prl \textbf{89}, 137903 (2002).

\bibitem {BS02}
    S. D. Bartlett and B. C. Sanders, ``Efficient classical simulation of optical quantum information
    circuits'', \prl \textbf{89}, 207903 (2002).

\bibitem {BSBN02}
    S. D. Bartlett, B. C. Sanders, S. L. Braunstein, and K. Nemoto, ``Efficient classical simulation of
    continuous variable quantum information processes'', \prl \textbf{88}, 097904
    (2002).

\bibitem {BSS89}L.~Blum, M.~Shub, and S.~Smale, ``On a theory of
    computation and complexity over the real numbers: NP-completeness, recursive
    functions and universal machines'', Bull. Am. Math.
    Soc. \textbf{21}, 1 (1989).

\bibitem {MP} ``Continuous turing machines'', Manuscript, not dated. (Available at \newline {
    http://www.mathpages.com/home/kmath135.htm}. Reference downloaded
    03 Dec. 2008.)

\bibitem {DJ92}
    D.~Deutsch and R. Jozsa,
    ``Rapid solution of problems by quantum computation'',
    Proc. Royal Soc. (Lond.) A \textbf{439}, 553 (1992).

\bibitem {DD85}
    D.~Deutsch,``Quantum theory, the Church-Turing principle and the universal quantum computer'',
    Proc. of the Royal Society (London) A \textbf{400}, 97 (1985).

\bibitem {CEMM98}
    R.~Cleve, A.~Ekert, C.~Macchiavello, and M.~Mosca,
    ``Quantum algorithms revisited'', Proc. Royal Soc. (Lond.) A \textbf{454}, 339-354 (1998).

\bibitem {Ba05}
    D.~Bacon, ``Lecture 4: Quantum algorithms from summer school in Siena,
    Italy. August 30-September 2, 2005'',  Manuscript, 2005 .(Available at \newline {
    http://www.cs.washington.edu/homes \newline
    /dabacon/teaching/siena}. Reference downloaded 20 Oct. 2008.)

\bibitem {BP02}
    S. L. Braunstein and A. K. Pati,
    ``Deutsch-Jozsa algorithm for continuous variables'', in~\cite{BP03}.

\bibitem {Gro96}
    L.~K.~Grover, ``A fast quantum mechanical algorithm for database search'',
    Proc.\ 28th Ann.\ ACM Symp. on Theory of Computing (STOC '96),
    pp. 212--219 (1996).

\bibitem {AIKRY04}
    A. Ambainis, K. Iwama, A. Kawachi, R. Raymond, and S. Yamashita,
    ``Robust quantum algorithms for oracle identification'', arXiv: quant-ph/0411204.

\bibitem {dlM05}
    R.~de~la~Madrid,
    ``The role of the rigged Hilbert space in quantum mechanics'',
    Eur. J. Phys. \textbf{26}, 287 (2005).

\bibitem {BCD02}
    R. ~Blume-Kohout, C. ~M. ~Caves, and I. ~H. ~Deutsch,
    ``Climbing mount scalable: {Physical} resource requirements for a scalable quantum
    computer'', Found. Phys. \textbf{32}, 1641 (2002).

\bibitem {Cav02}
    C.~M.~Caves,
    ``Physical resources, entanglement, and the power of quantum
    computation'', Lecture given in SQuInT Summer
    Retreat University of Southern California, 2005 July 7. Reference downloaded 03 Dec. 2008.

\bibitem {Tan}
    C.~I.~Tan,
    ``Notes on Hilbert Space'', Manuscript, not dated. (Available at \newline {
http://jcbmac.chem.brown.edu/baird/QuantumPDF}. Reference
downloaded 20 Oct. 2008.)

\bibitem {BdGS02}
    S. D. Bartlett, H. de Guise, and B. C. Sanders, ``Quantum encodings in spin systems and harmonic oscillators'',
    \pra \textbf{65}, 052316 (2002).

\bibitem {Bra86}
    R.~N.~Bracewell,
    \emph{The Fourier Transform and Its Applications, 2nd Ed.}
    (McGraw-Hill, New York, 1986).

\bibitem {Ad04}
    M.~Adcock
    ``The classical and quantum complexity of the Goldreich-Levin problem
    with applications to bit commitment'',
    \emph{Master of Science Thesis, Department of Computer Science, University of
    Calgary (2004)}.

\bibitem {Can01}
    J.~F.~Canny, ``Chernoff bounds'', Manuscript, 2001.
    (Available at http://www.cs.berkeley.edu/~jfc/cs174/lecs /lec10/lec10.pdf. Reference downloaded 20 Oct. 2008.)

\bibitem {CLR90}
    T.~H.~Cormen, C.~E.~Leiserson, and R.~L.~Rivest, \emph{Introduction to Algorithms},
    MIT Press, Cambridge, MA 1990

\bibitem {HS05}
    P.~H\o yer and R.~\v{S}palek ,
     ``Lower Bounds on quantum query complexity'',
     Bull. Euro. Assoc. for Theoretical Computer Science Vol. \textbf{87}, 2005.

\bibitem {Per86}
    A.~Perelomov, \emph{Generalized Coherent States and Their Applications}
    (Springer-Verlag, New York, 1986).

\bibitem {ACGT72}
    F.~T.~Arrechi, E.~Courtens, R.~Gilmore, and H.~Thomas, ``Atomic coherent states in quantum optics'',
    \pra \textbf{6}, 2211 (1972).

\bibitem {KU93}
    M.~Kitagawa and M.~Ueda, ``Squeezed spin states'',
    \pra \textbf{47}, 5138 (1993).

\bibitem {Le97}
    U.~Leonhardt, \emph{Measuring the Quantum State of Light}
    (Cambridge University Press, Cambridge UK, 1997).

\bibitem {BV97}
    E.~Bernstein and U.~V.~Vazirani, ``Quantum complexity theory'', SIAM J. on Comp.
    \textbf{26}, No.~5, pp.~1411--1473 (1997).

\bibitem {CvDNT99}
    R.~Cleve, W.~van~Dam, M.~Nielsen, and A.~Tapp, ``Quantum entanglement and the communication complexity of
    the inner product function'', Lecture Notes in Computer Science \textbf{1509}
    (Springer-Verlag), pp.\ 61-74 (1999).

\bibitem {Sim97}
    D.~R.~Simon,
    ``On the power of quantum computing'',
    SIAM J. on Comp. \textbf{26}, No.~5, pp.~1474--1483 (1997).

\bibitem {Sho97}
    P.~W.~Shor,
    ``Polynomial-time algorithms for prime factorization and discrete logarithms on a quantum computer'',
    SIAM J. on Comp. \textbf{26}, No.~5, pp.~1484--1509 (1997).

\bibitem {AC02}
    M.~Adcock and R.~Cleve,
    ``A quantum Goldreich-Levin theorem with cryptographic
    applications''
    \emph{Proc. 19$^\text{th}$ International Symp. Theor. Aspects Comp. Sci. (STACS 2002)},
    H. Alt and A. Ferreira, eds., Lecture Notes in Computer Science \textbf{2285}
    (Springer-Verlag), pp.\ 323-334 (2002).

\end{thebibliography}
\end{document}